\documentclass[twocolumn,tighten]{aastex701}
\usepackage[T1]{fontenc}
\usepackage{apjfonts}
\usepackage[figure,figure*]{hypcap}
\usepackage{amsmath}
\usepackage[capitalise,noabbrev]{cleveref}
\usepackage{graphicx}
\usepackage{multirow}

\begin{document}

\title{2024 YR$_4$: Identification of Possible Precoveries in 2016 IPTF Data}

\author[0009-0004-6814-5449]{Sam Deen}
\affil{Independent Researcher}
\affil{Deep Random Survey, deeprandomsurvey.org}
\email{planetaryscience@yahoo.com}

\author[0009-0002-5890-6289]{Derek Lam}
\affil{Independent Researcher}
\email{derek@lam.io}

\begin{abstract}

2024 YR$_4$ is a 40-100 meter-diameter asteroid and former Torino Scale 3 object which currently has a roughly 4\% chance of impacting the Moon on 2032 December 22, an event which recent studies suggest could pose a hazard on Earth due to impact ejecta. We present a search for, and identification of, potential precovery observations of the virtual lunar impactor in Intermediate Palomar Transient Facility (IPTF) survey data, as well as other publicly accessible surveys, dating from 2016. These candidate detections, not accounting for any currently-undetected Yarkovsky forces, predict a perilune of $22001 \pm 49$ km and a perigee of $277534 \pm 46$ km (relative to the center of each respective body) representing an improvement of $>300$ times in the approach distance uncertainty above the existing orbit solution and, if confirmed, decisively ruling out a lunar impact in 2032. Using a matched filter tuned to 2024 YR$_4$'s predicted appearance in each image, we find the detection to be significant at $P_{null}\approx 5 \times10^{-9}$. The resultant possible orbit solution should be easy to confirm during 2024 YR$_4$'s 2028 approach to Earth, potentially greatly reducing the effort required by the planetary defense community at large to characterize 2024 YR$_4$ before its potential lunar impact.

\end{abstract}

\section{Introduction}\label{sec:intro}

Since the inception of modern asteroid surveys shortly before the turn of the millennium, planetary defense has always been a priority among minor planet researchers. The discovery, physical characterization, and prediction of possible impacts for asteroids on Earth-crossing orbits has been routinely carried out by countless surveys of various designs ever since, with computational and organizational support for tracking potential future Earth-impacting asteroids (termed 'virtual impactors') provided by the NASA's Center for Near-Earth Object Studies (CNEOS)\footnote{https://cneos.jpl.nasa.gov/sentry/intro.html}, ESA's Near-Earth Objects Coordination Center (NEOCC)\footnote{https://neo.ssa.esa.int/risk-list} and Near Earth Objects Dynamic Site (NEODyS)\footnote{https://newton.spacedys.com/neodys/index.php?pc=4.0}, as well as numerous others.

To this end, a handful of methodologies have been devised to indicate the relative risk of an asteroid's uncertain future impacts, and the associated priority at which it should be studied. One of these, the Torino Scale (\citealt{Binzel1997}, \citealt{Binzel2000}), grades virtual impactors on a semi-arbitrary scale of 0-10 using the probability of a future impact combined with the asteroid's estimated size. The vast majority of virtual impactors as tracked by these groups have never ranked higher than 0 on the scale, which entails either an estimated asteroid diameter of under 20 meters regardless of other circumstances, or an increasingly low probability of impact (starting at $<1$ in 1000) for proportionally larger diameters, and this value is intended to signify a very low priority for scientific and public attention. Torino Scale 1, for comparison, signifies a similarly low priority but is comparatively infrequent, with only a handful of asteroids achieving this ranking annually as their orbital uncertainty-based line of variations (LOV) briefly intersects Earth with slightly more probability before being invariably reduced to Torino 0 or taken off the list entirely. As of the writing of this paper, the most recent asteroid to reach Torino 1 was 2025 FA$_{22}$, a hundred-meter-scale asteroid which initially reached a 1-in-8300 (0.012\%) impact probability before further observations ruled out an impact entirely.

Torino ratings above 1 are increasingly rare, and prior to 2024 had only been briefly achieved by (144898) 2004 VD$_{17}$ for three months in 2006 (Torino 2) and 99942 Apophis during a single week in 2004 (Torino 4). While neither of these asteroids ultimately posed an absolutely high impact risk to Earth, their rarity has made it especially prudent to prioritize followup studies, both as practice for predicting an eventual genuine impact event, and to rule these objects out as such an impactor themselves.

2024 YR$_4$ is a 40-100 meter-diameter asteroid discovered on 2024 December 27 by the ATLAS survey's Chile telescope (W68) which was quickly identified within a week as having a credible impact threat to Earth on 2032 December 22. Initial followup observations in the ensuing month continued to increase the probability as its line of variations converged closer to Earth, and by 2025 Jan 27 it reached Torino Scale 3, signifying an Earth impact chance of greater than 1\%. Extensive followup observations from many facilities resulted in the impact probability briefly peaking at 3.1\% (1 in 32) before later tracking, largely led by the James Webb Space Telescope (JWST) in 2025 March through May, ultimately ruled out the 2032 impact with Earth, with the most recent orbital solution (JPL 78) giving an approach distance of $270000\pm 69000$ km.

This followup has not, however, ruled out an impact in December 2032. Although not officially tracked by agencies, a lunar impact remains credible with the current orbital uncertainty, with the aforementioned most recent orbital solution also giving a lunicentric approach distance of $10682 \pm ^{74001} _{9667}$ km, allowing for a perilune of as little as 1015 km compared to the Moon's mean radius of 1737 km. As of writing, there remains a roughly 4\% chance of lunar impact, a possibility which recent analysis by \citep{Wiegert2025} found could still pose a tangible hazard to Earth due to impact ejecta.

As such, despite an Earth impact having been ruled out, efforts to constrain 2024 YR$_4$'s orbit are ongoing, but are delayed by the asteroid's highly eccentric orbit and faint absolute magnitude which render it essentially invisible for most of its four-year orbit. As a result, followup observations of 2024 YR$_4$ will not be practically possible until at least 2028, only a single orbit before its potential lunar impact and inappropriately late to prepare any meaningful planetary defense response such as an asteroid redirect mission if an impact were to be confirmed. As such, early on many astronomers looked to potential pre-discovery observations ("precoveries") taken by various survey telescopes in the years prior to 2024 YR$_4$'s discovery, which may allow us to achieve a multi-year observation arc and fully confirm or rule out the asteroid's 2032 impact without needing to wait several years to do so.

In this paper, we present the results of our searches for precoveries of 2024 YR$_4$, in which we find a compelling candidate in a series of images taken by the Intermediate Palomar Transient Factory (IPTF) asteroid survey in 2016. In Section \ref{sec:apparitions} we will present 2024 YR$_4$'s past observing circumstances and identify several potential precovery images. In Section \ref{sec:analysis} we will describe our methods and analyze the identified imagery. In Section \ref{sec:candidate} we will review a promising candidate detected in the IPTF dataset in concert with the other available data, with a more rigorous analysis made in Section \ref{sec:significance}. Finally in Section \ref{sec:summary} we will summarize the results of the precovery search and discuss the predictions made by these results which should be readily verifiable during 2024 YR$_4$'s 2028 approach.

\section{Precovery Identification}\label{sec:apparitions}

At present, 2024 YR$_4$ has an eccentric, almost exactly 4-year orbit with a low ecliptic inclination of just 3.4 degrees. This, combined with its close approach (perigee 2.156 Lunar distances (LD)) in 2024, results in it having close approaches to Earth every four years, resulting in the aforementioned recovery opportunity in 2028 (perigee $20.84 \pm 0.12$ LD) and previously possible impact in 2032 (perigee $0.87 \pm 0.18$ LD). However, prior to being perturbed onto this near-resonant orbit in 2024 it had a slightly longer 4.05-year period resulting in a much more distant approach in 2020 (perigee $54.182 \pm 0.022$ LD) and a close approach in 2016 on the incoming leg of its orbit rather than the outgoing leg as seen in all later encounters (perigee $29.59 \pm 0.18$ LD), the latter of which was 2024 YR$_4$'s closest approach to Earth since 1964. Further details of recent encounters are provided in Table \ref{tab:YR4 approaches}. 

\begin{deluxetable}{ccccc}
\label{tab:YR4 approaches}
\tablecaption{Near-past and near-future Earth encounter details}
\tablehead{
\colhead{Date\tablenotemark{a}} & \colhead{approach dist} & \colhead{peak mag} & \colhead{elong\tablenotemark{b}} & \colhead{3$\sigma$ ellipse\tablenotemark{b,}\tablenotemark{c}} \\ & (au) & (mag, V) & (deg) & (arcsec)}
\startdata
2012-08-29 & $0.19546$ & 22.90 & 109 & $1084 * 0.064$ \\
2016-09-08 & $0.07603$ & 20.64 & 125 & $2170 * 0.258$ \\
2020-10-26 & $0.13922$ & 23.47 & 103 & $83.5 * 0.127$ \\
2024-12-25 & $0.00554$ & 15.83 & 106 & $1.90 * 0.243$ \\
2028-12-17 & $0.05355$ & 20.79 & 105 & $37.1 * 0.185$ \\
2032-12-22 & $0.00178$ & 13.39 & 105 & $2600 * 1.18$ \\
\enddata
\tablenotetext{b}{at closest approach}
\tablenotetext{b}{at peak magnitude}
\tablenotetext{c}{major and minor axis dimensions of uncertainty ellipse}
\end{deluxetable}

Due to the poor geometry of most of the pre-2024 approaches, only the 2016 approach reached a magnitude accessible to most contemporary surveys. A search using the Canada Astronomical Data Center's Solar System Object Image Search (SSOIS)\footnote{https://www.cadc-ccda.hia-iha.nrc-cnrc.gc.ca/en/ssois/} \citep{Gwyn2012} similarly returned no compelling archival data from either 2012 or 2020, but several images were identified as possibly containing 2024 YR$_4$ in 2016. An additional search for archival Palomar Transient Facility (PTF) imagery\footnote{https://irsa.ipac.caltech.edu/applications/ptf/} in the NASA/IPAC Infrared Science Archive (IRSA) turned up several more images taken as part of the Intermediate PTF survey (IPTF). Additional searches using the Planetary Data System's Small Bodies Node (PDS SBN) CATCH tool\footnote{https://catch.astro.umd.edu/} as well as EURONEAR's Mega Precovery tool\footnote{https://s141.central.ucv.ro/tools/megaprecorb.php} did not return any additional images. The collected candidate images are described in Table \ref{tab:possible images}.

\begin{deluxetable}{ccccccc}
\label{tab:possible images}
\tablecaption{Candidate archival images}
\tablehead{
\colhead{Date} & \colhead{Telescope} & \colhead{exposures} & \colhead{filter} & \colhead{exp. mag} & \colhead{3$\sigma$ ellipse\tablenotemark{a}} & \colhead{sky motion} \\ & & (sec) & & (mag, V) & (arcsec) & (arcsec/min)}
\startdata
2016-08-02 12:55-13:25 & Subaru 8.3m/HSC   & 5x420s & NB921      & 23.52 & 232*0.023  & 1.026  \\
2016-08-09 09:54       & Subaru 8.3m/HSC   & 200s   & Y          & 22.89 & 529*0.033  & 1.392  \\
2016-08-11 05:01       & CTIO 4m/DECam     & 107s   & z          & 22.72 & 569*0.037  & 1.534  \\
2016-08-25 07:32-09:45 & Palomar 1.2m/IPTF & 4x60s  & r, g, g, r & 21.31 & 1171*0.101 & 4.528  \\
2016-08-28 09:18-10:13 & Palomar 1.2m/IPTF & 2x60s  & g, r       & 21.02 & 1427*0.134 & 6.242  \\
2016-08-31 09:44-10:40 & Palomar 1.2m/IPTF & 2x60s  & g, r       & 20.78 & 1756*0.181 & 8.798  \\
2016-09-03 09:49       & Palomar 1.2m/IPTF & 60s    & g          & 20.65 & 2116*0.246 & 12.345 \\
\enddata
\tablenotetext{a}{major and minor axis dimensions of uncertainty ellipse}
\end{deluxetable}

The first series of images, taken by the Hyper-SuprimeCam (HSC) instrument on the 8.3m-aperture Subaru telescope on Maunakea, were taken in a narrowband filter with effective wavelength 918 nm and an effective width of 14 nm, roughly between the instrument's z and Y filters. While a total of 35 minutes of time was integrated near 2024 YR$_4$'s location, the asteroid's predicted motion of 7.2 arcsec in each exposure combined with the narrowband filter would make this a challenging detection despite the telescope's large aperture. Similarly, the single Y-band exposure on August 9 suffers from HSC's low quantum efficiency (generally <40\%) at the filter's long wavelengths, although the brightening asteroid and much wider filter may make a detection comparatively easier.

\section{Analysis of Images}\label{sec:analysis}

Each of the images identified in the previous section was investigated by eye using the Strasbourg Astronomical Data Center's (CDS) Aladin software\footnote{https://aladin.cds.unistra.fr/} and compared with other images taken by the same survey of each location to identify any transients present in the expected image but not the other. Upon finding a marginal detection in any image, we then generated a provisional orbit solution using that point as a measurement, checking the other candidate imagery for detections at the now-expected position given the orbit solution.

Despite the large major axis of 2024 YR$_4$'s uncertainty ellipse (as much as 35 arcminutes in the 09/03 image) the minor axis is consistently extremely small (<0.25 arcsec at every date) rendering the asteroid's line of variations (LOV) as effectively one dimensional on each date and severely restricting the area in which a marginal detection would be credible.

At the time we performed this search from January through April 2025, the uncertainty region was much longer, largely on account of the JWST observations of March and May 2025 being unreported at the time. As a result, roughly three times the displayed LOV was searched, but will not be reflected here due to a lack of promising candidates in this outer region and the currently much-improved knowledge of 2024 YR$_4$'s location on these dates.

\begin{figure*}
\centering
\includegraphics[width=\textwidth,angle=0]{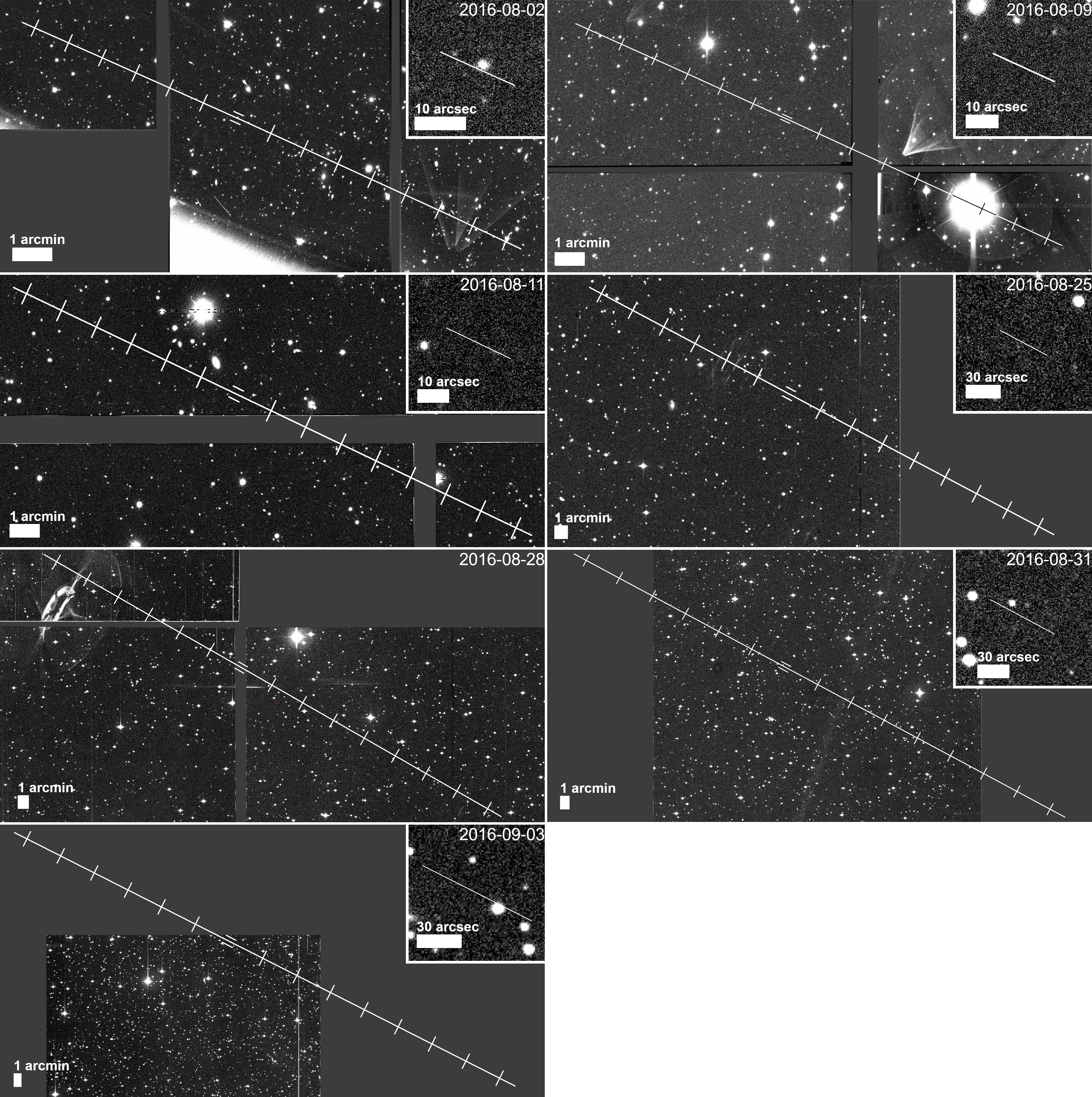}
\caption{(Left to right, top to bottom) Representative frames of the datasets searched for 2024 YR$_4$ as described in Table \ref{tab:possible images}, in chronological order. The 3$\sigma$ line of variations is indicated with a line with the associated 2032 perilune distances marked with ticks at 10,000 km intervals, and positions corresponding to a 2032 lunar impact indicated with parallel lines -- displayed in greater detail on the upper right inset for each image. North is up and East is left; the true thickness of the LOV is less than a pixel in each image, including the insets.
\label{fig:Candidate_images}}
\end{figure*}

The 3-$\sigma$ uncertainty region is overlaid on a sample image from each of the eight nights in Figure \ref{fig:Candidate_images}. Note that the August 2 images were continuously dithered by a few arcminutes between each exposure, and the asteroid's motion between consecutive IPTF exposures on the same night was several arcminutes, so the chip gaps on these particular nights should not be taken as representational of all of the images on that night. Also note that the length of the LOV inflates by roughly an order of magnitude between the first and last images as the asteroid approached Earth.

While the point-source limiting magnitude on 08/02 of each individual exposure was approximately z=23.3, the strong expected motion streaking of 2024 YR$_4$ in each image (spanning 43 pixels, roughly 9x the average stellar FWHM) diminishes this sensitivity significantly, making it potentially impossible to detect the V=23.5 2024 YR$_4$ on this date. The 08/09 image is deeper with a limiting magnitude of Y=23.5, with less streaking (only 28 pixels, roughly 4.5x the stellar FWHM) but nonetheless fails to appear as a significant detection. The LOV on this date is partially contaminated with the bright Y=6.0 red giant star HD 212849, complicating transient detection from positions corresponding to perilunes of 51000 to 65000 km. A pair of chip gaps between positions corresponding to perilunes of 20000 to 28000 km and 31000 to 37000 km are also present.

The 08/11 DECam image, despite a deep limiting magnitude of z=22.7, fails to detect any transients consistently appearing in other data, although notably a pair of CCD chip gaps exist extending from positions corresponding to 2032 perilunes of 13000 to 29000 km, and 51000 to 57000 km. These overlap significantly with the aforementioned blind spots on 08/09, indirectly providing evidence that 2024 YR$_4$ will achieve a perilune of somewhere between either 20000 and 28000 km, or between 51000 and 57000 km, through negative observation. In either case, the images from 08/09 and 08/11 cover roughly 80\% of 2024 YR$_4$'s LOV at least once with sufficient depth to detect it if it were in these locations, with both covering the entire virtual lunar impact region, largely discounting the possibility of a lunar impact regardless of where 2024 YR$_4$ itself may be.

\section{IPTF Candidate Detection}\label{sec:candidate}

\begin{deluxetable}{ccccccc}
\label{tab:IPTF images}
\tablecaption{Details of each IPTF exposure}
\tablehead{
\colhead{Date} & \colhead{filter} & \colhead{FWHM\tablenotemark{a}} & \colhead{lim. mag.\tablenotemark{b}} & \colhead{YR$_4$ mag} & \colhead{YR$_4$ SNR} & \colhead{LOV covered}
\\ & & (arcsec) & (mag, 3$\sigma$) & (mag, in filter) & & (\%)
}
\startdata
08-25 07:32:39.2 & r & 2.52 & 21.36 & 21.20 & 3.48 & 54.8 \\
08-25 08:24:10.9 & g & 3.00 & 21.00 & 21.69 & 1.59 & 52.0 \\
08-25 08:49:30.2 & g & 4.31 & 21.01 & 21.69 & 1.60 & 66.0 \\
08-25 09:45:40.9 & r & 4.32 & 21.14 & 21.20 & 2.84 & 69.7 \\
08-28 09:18:23.4 & g & 2.82 & 21.28 & 21.41 & 2.66 & 95.0 \\
08-28 10:13:27.3 & r & 2.45 & 21.20 & 20.91 & 3.92 & 94.7 \\
08-31 09:44:25.0 & g & 3.26 & 21.06 & 21.16 & 2.74 & 66.5 \\
08-31 10:40:15.8 & r & 3.21 & 21.15 & 20.67 & 4.67 & 66.2 \\
09-03 09:49:29.1 & g & 2.70 & 21.29 & 21.03 & 3.81 & 20.1 \\
\enddata
\tablenotetext{a}{Median FWHM parameter in image header}
\tablenotetext{b}{3$\sigma$ limiting magnitude for the streaked 2024 YR$_4$, based on the limiting magnitude for point sources in each image as assessed by the Tycho software \citep{Parrott2020} -- using the ATLAS All-Sky Stellar Reference Catalog \citep{tonry2018} -- and the amount of predicted streaking in that respective image. The SNR listed in the adjacent column is calculated using this limiting magnitude.}
\end{deluxetable}

The Intermediate Palomar Transient Factory (IPTF) operated the 1.2-meter Samuel Oschin Telescope at Mount Palomar between 2012 and 2017 as an intermediate survey between the Palomar Transient Factory (PTF; 2009-2012) and the ongoing Zwicky Transient Facility (ZTF; 2018-present) with a largely similar survey design to IPTF, although unlike in PTF only some of the survey's images have been publicly released to date. Despite this, IPTF's pointings are nonetheless queryable on IRSA through the same portal as PTF. With the assistance of Frank Masci on behalf of the IPTF team we were granted access to the unreleased images covering 2024 YR$_4$'s nominal location listed in Table \ref{tab:possible images}. Each of these images, split between r- and g-filter on a given night, has 3$\sigma$ point source limiting magnitudes in the range r=21.2-21.5, g=21.1-21.7 mag, although in practice these will be lower for the strongly streaked 2024 YR$_4$ depending on the date; the lower values and other details about each exposure are outlined in Table \ref{tab:IPTF images}.

As they were acquired with the same instrument, a near-identical pixel scale, and consistent filters, ZTF Deep Reference images were acquired from IRSA to cover the full LOV at each image epoch for comparison to the IPTF data. Due to being stacks of several hundred exposures taken over the course ZTF's ongoing several-year survey, these images are considerably deeper than any of the IPTF exposures, with the archive listing 5$\sigma$ limiting magnitudes of fainter than magnitude 23 in both filters for most of the fields used. This allows marginal signals in IPTF corresponding to stars to be obvious in the reference images.

Each of the images was then checked against the reference data for marginal detections near the LOV which did not correspond to any significant detections in the reference imagery. For each resultant signal, we recalculated a provisional orbit for 2024 YR$_4$ using this position and searched for additional marginal signals corresponding to that orbit in the remaining IPTF, DECam, and Subaru images where available. A genuine signal, however marginal, should consistently appear across most images of similar depth, while a spurious signal may appear significant in one or two images but be insignificant in most.

\begin{figure}
\centering
\label{fig:rawdetections}
\includegraphics[scale=0.7,angle=0]{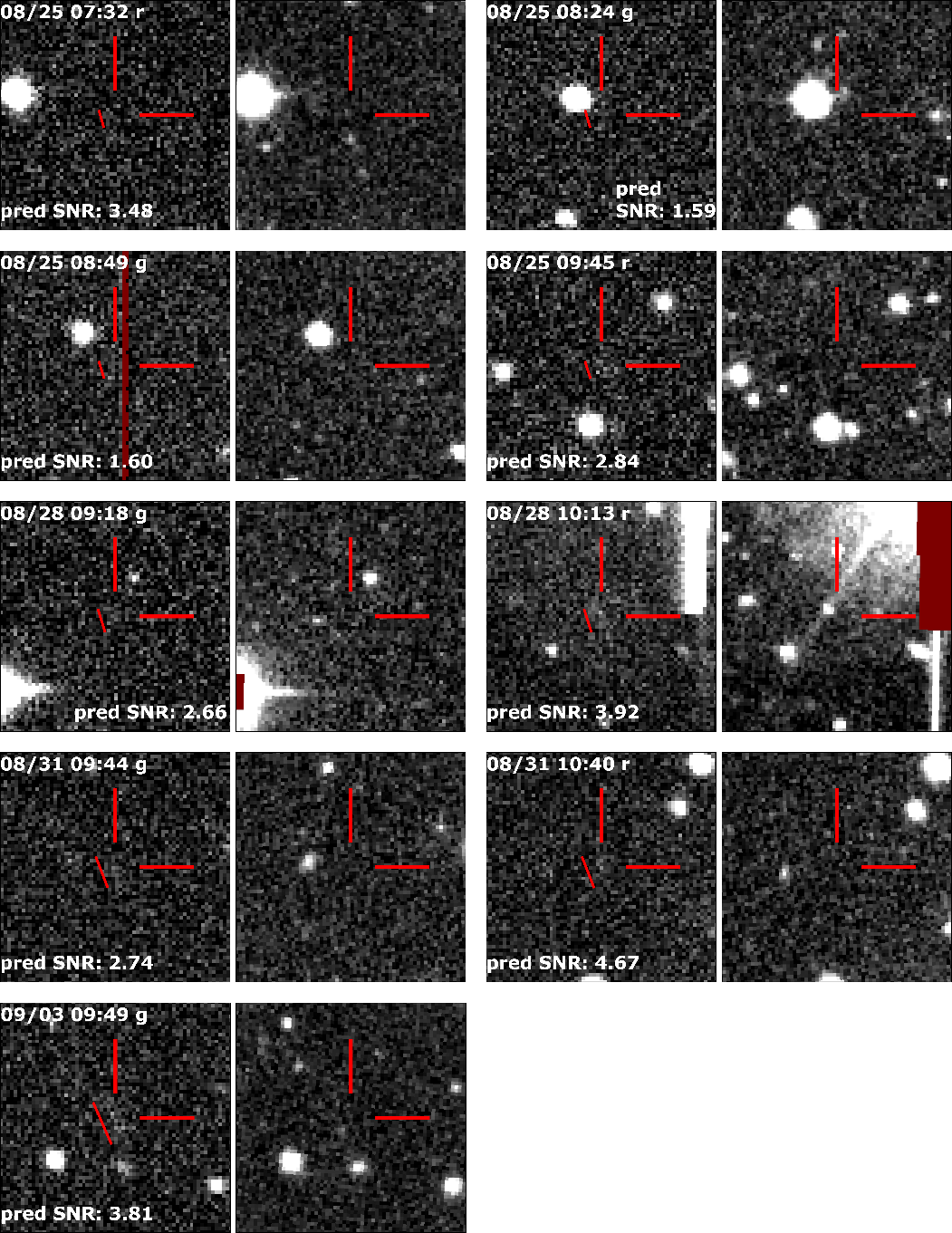}
\caption{(Left of each column) Detections of a potential candidate for 2024 YR$_4$ in each individual IPTF image, with (right of each column) ZTF deep reference images in the same filter for comparison, centered on the expected location on each date using the nominal orbit produced by these measurements. The direction and length of 2024 YR$_4$'s motion streaking over the course of each exposure is indicated with a line. The expected SNR as described in Table \ref{tab:IPTF images} is included as well.
\label{fig:Rawdetections}}
\end{figure}

Although a handful of single-image marginal detections appeared in the process of searching, ultimately the density of IPTF survey images on these dates made it straightforward to compare and rule out many of the insignificant signals and/or artifacts. Although a large cross section through these images much larger than the current 3-sigma LOV was originally investigated, only a single signal appeared consistently between all of the images. Figure \ref{fig:rawdetections} depicts the detections in each respective image compared against the aforementioned ZTF deep reference images. The expected motion streak in each image is also included, and shows good agreement with the shape of each detection, especially towards the latter imagery where the asteroid is predicted to be brighter and more strongly streaked.

\section{Significance Analysis}\label{sec:significance}

\begin{figure*}
\centering
\includegraphics[width=\textwidth]{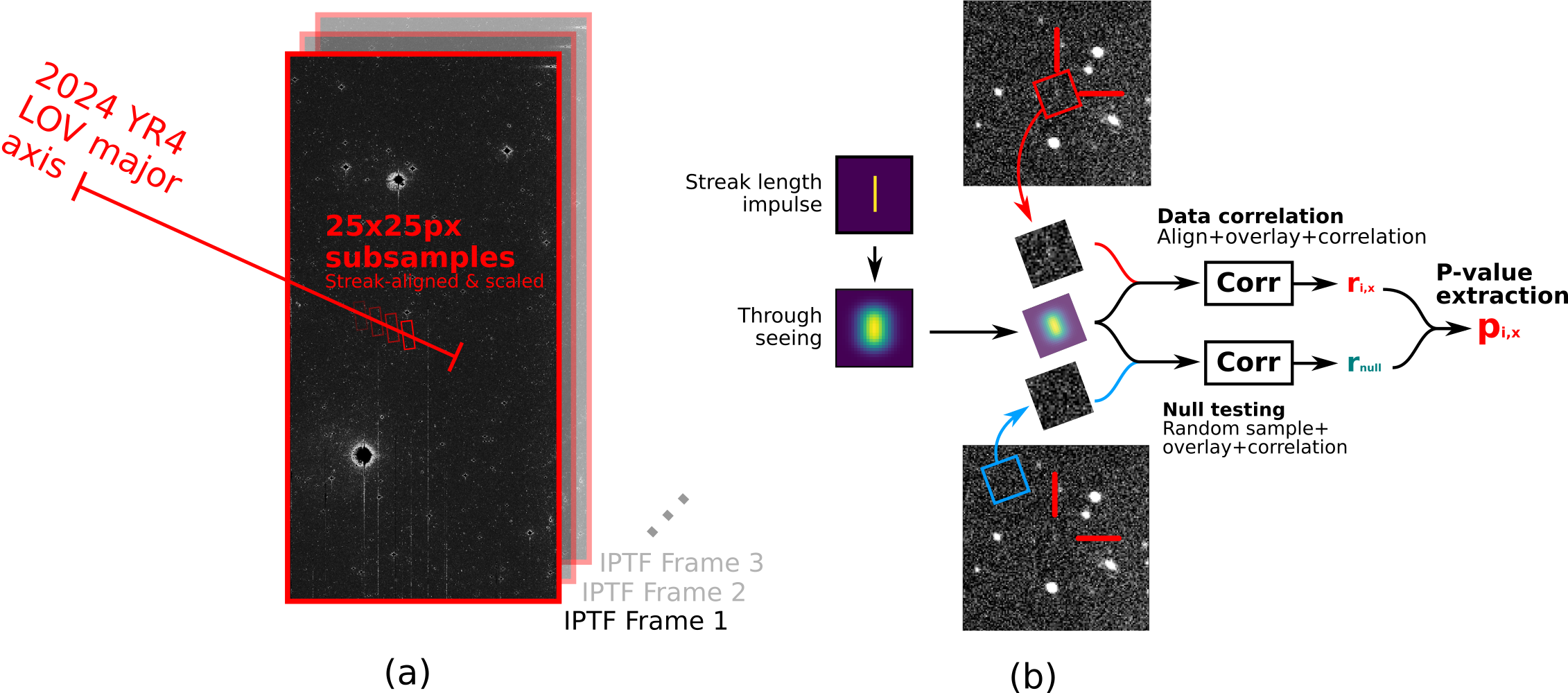}
\caption{Automated detection and significance testing methodology. $N$ samples were gathered in \textbf{(a)} along the LOV, then corrected with a synthetic flat. Reference images from ZTF were aligned, matched to the IPTF seeing, gathered along the LOV, flat-corrected and subtracted from the IPTF samples. Synthetic images of 2024 YR$_4$ were generated in \textbf{(b)} matching the streak length, seeing and expected magnitude of 2024 YR$_4$ at the time each IPTF frame was taken. The compensated image samples were then correlated against these reference images in \textbf{(b)}, and null correlations were performed against the rest of the image \textbf{(b)} to estimate P-value $p_{i,k}$ for the data correlation coefficients $r_{i,k}$. Null and true detection distributions were also calculated from these same images through simulation of overlaid streaks in non-LOV areas. These null and true distributions were used to implement a maximum likelihood test against the P-values calculated in \textbf{(b)} to extract the most likely detection candidate[s], as shown in \cref{fig:detection-testing}}
\label{fig:methodology-flowchart}
\end{figure*}

The detection was assessed using an automated search along the LOV major axis. This methodology is illustrated in \cref{fig:methodology-flowchart}.

First, we performed a starfield subtraction of the IPTF frames using the ZTF deep reference images described in Section \ref{sec:candidate} with linear brightness fitting. Reference frames were Gaussian blurred to empirically match the reference images' FWHM to each IPTF image's FWHM. Then, 2000 subimages were sampled uniformly along the LOV major axis from each frame of index $i$, resulting in samples spaced at approximately the seeing FWHM in each frame. These subimages are indexed by their position along the LOV major axis by the LOV sample ``coordinate'' $x\in[0,1]_{N}$ where $N$ represents the number of subdivisions of the LOV interval. All image alignment and coordinate manipulation were performed with Astropy \citep{astropy:2022}.

Dimension and orientation of the subimages were standardized between all frames, to a width of $3FWHM$, a height of $2l_{streak}$, and an orientation where the height of the image is parallel to 2024 YR$_4$'s motion in the frame. These subimages were constructed from the frame pixels using a 2D piecewise linear interpolator, mapping the $3\cdot FWHM\times 2l_{streak}$ rectangle in RA-DEC frame space to a standardized subimage size of 25px$\times$25px, in order to fully resolve the longest streak ($l_{streak}=12.345''$ at 2016-09-03 09:49, see Table \ref{tab:possible images}).

We generated reference images of 2024 YR$_4$ based on its expected motion over each exposure, its expected magnitude, and the seeing of each frame. First, we applied the seeing of each frame to a vertical line one pixel wide at high resolution ($>500$px$\times 500$px), to approximate 2024 YR$_4$'s streaked impulse response (\cref{fig:methodology-flowchart}(b)). This image was aligned along the expected motion of 2024 YR$_4$ in each frame. These aligned reference images were then correlated to each subsample along the LOV to produce the Pearson correlation coefficients $r_{i,x}$, for each frame $i$ and each LOV sample coordinate $x$.

The empirical distribution of correlation coefficients was constructed by sampling 200 chords of the frame parallel to the LOV. This null distribution of correlations $r_{null}$ was fitted to a Gaussian with an separate piecewise tail model (due to residual dim brightness sources -- the significance of this upper non-Gaussian brightness tail prompted the use of this empirical reference distribution). Tails were isolated using an outlier filter similar to Grubbs's test \citep{Grubbs1950}, where all minimal and maximum values that deviated a normal CDF by more than a factor of $\alpha$ were iteratively removed and the distribution refitted until no deviating points remained.

\begin{figure*}
    \centering
    \includegraphics[width=\textwidth]{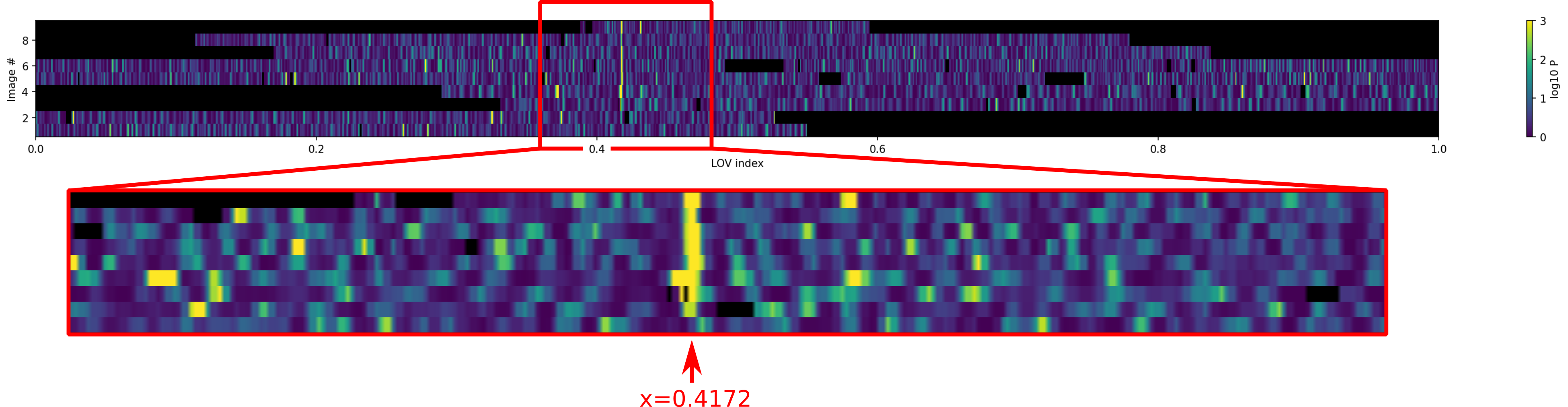}
    \caption{Log-P-values $-{\log{(p_{i,x})}}$ plotted against frame index ($i$, vertical) and LOV sample coordinate ($x$, horizontal). Brighter regions are more significant (larger $-{\log{(p)}}$). The candidate detection is highlighted by a red arrow in the inset.}
    \label{fig:P-vs-image-lov}
\end{figure*}

The resulting matrix of P-values $p_{i,x}$ as a function of frame index $i$ and LOV sample coordinate $x$ are plotted in \cref{fig:P-vs-image-lov}.

\begin{figure}
    \centering
    \includegraphics[width=\linewidth]{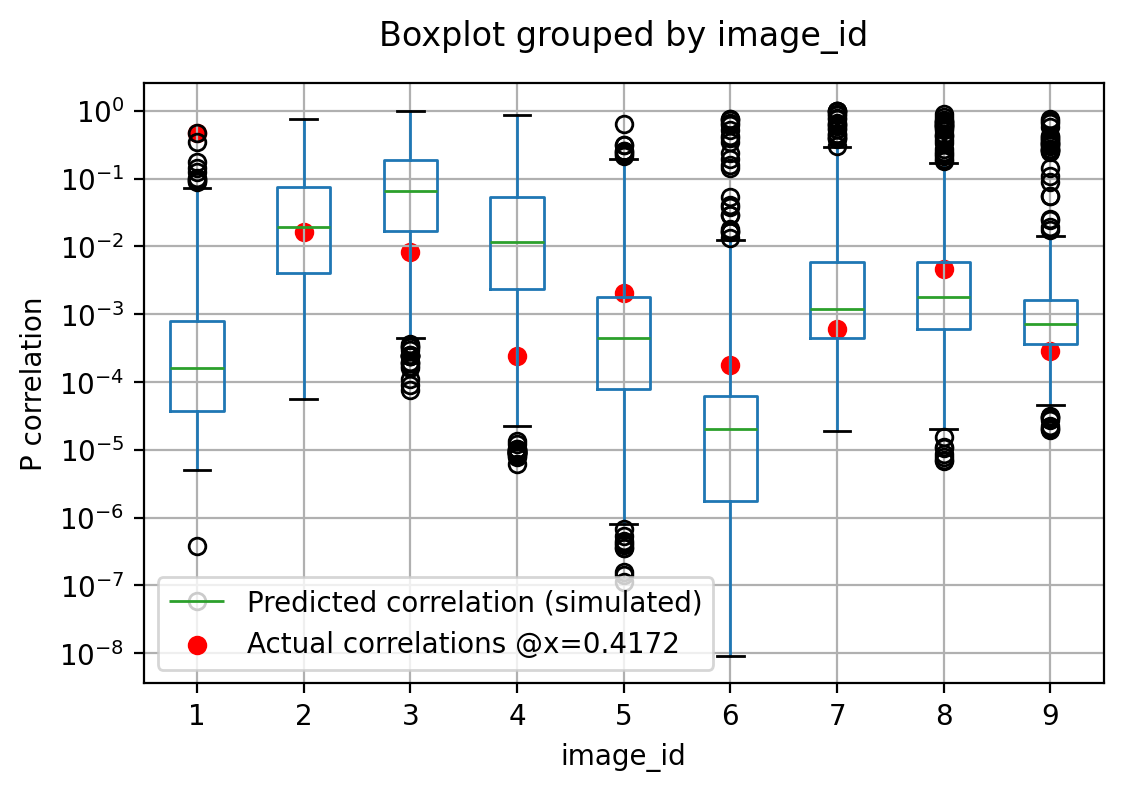}
    \caption{Boxplots of predicted correlation P-value for each IPTF frame from simulated 2024 YR$_4$ detections. The actual correlations from the most significant detection at $x=0.4172$ are indicated by red points.}
    \label{fig:correlation-boxplots}
\end{figure}

The P-values $p_{i,x}$ were then multiplied across frames $i$ to calculate the joint P-value of correlation coefficients for each LOV sample coordinate $x$ (\cref{eq:joint-p-calc}).

\begin{equation}
    P_x=\prod_{i=1}^{N_x} \left[ p_{i,x}\right]
    \label{eq:joint-p-calc}
\end{equation}

Invalid subimages -- where the LOV was outside of the frame, or had $> 50\%$ of the reference image covered by bad pixels -- were excluded. Therefore, the number of valid frames $N_x$ varies for each index $x$, accounting for excluded subimages. These joint P-values $P_x$ were then assessed by comparing to true and null distributions to determine the confidence of detections at each $x$. The following sections describe the derivations of these two distributions.

\begin{figure*}
\centering
\includegraphics[width=\textwidth]{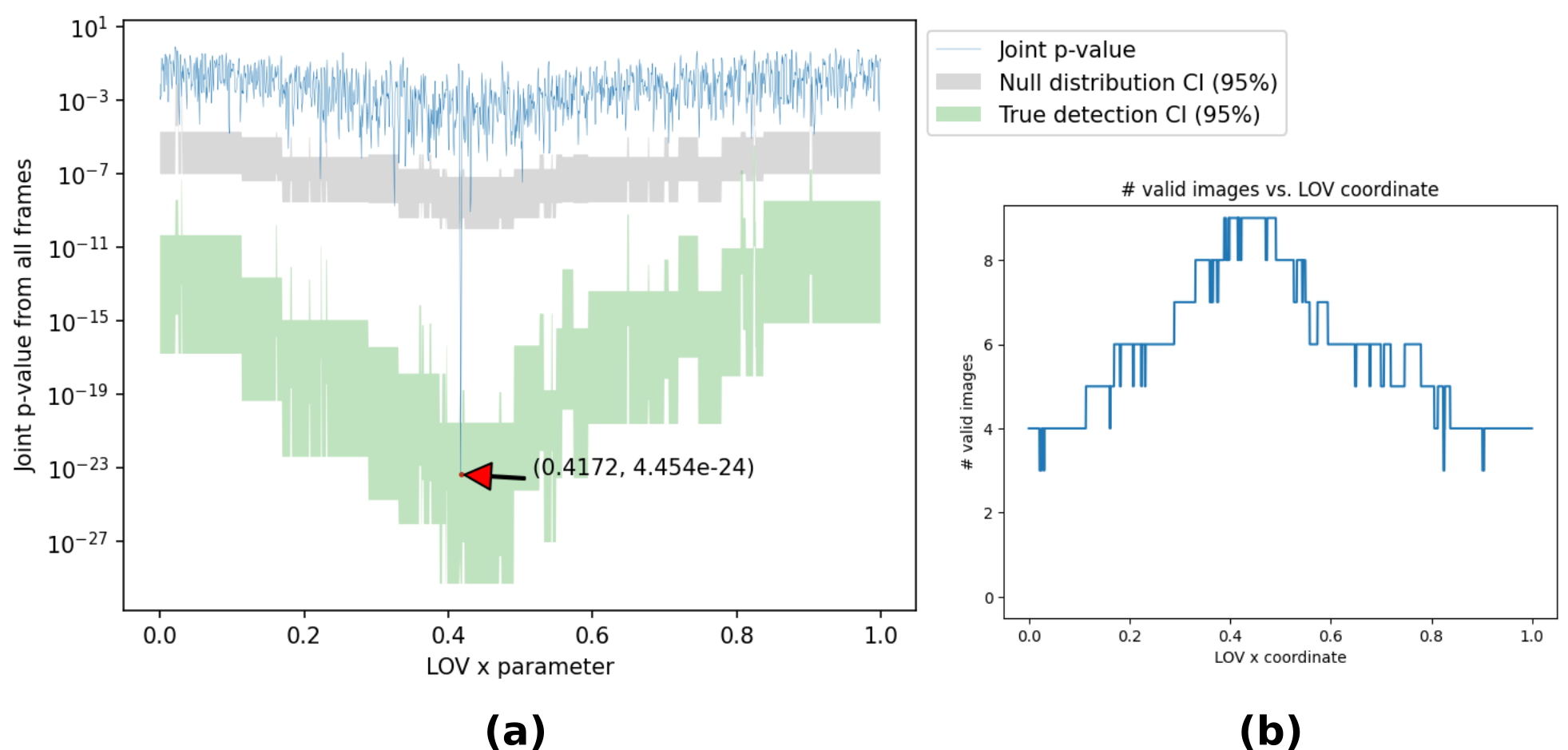}
\caption{\textbf{(a)} Joint P-values $P_x$ of 2000 samples along the LOV major axis. Most significant detection is denoted by the arrow, at $x=0.4172$. The 95\% confidence intervals for both the null and true detection distributions are shown by the grey and green bands respectively. The full null and true detection distributions are shown in \cref{fig:detection-testing} for this candidate. \textbf{(b)} \# Valid frames for each LOV sample coordinate. Represents the degree of coverage of each point along the LOV from IPTF frames.}
\label{fig:JointP}
\end{figure*}

\subsection{Null distribution}
\label{ssec:null-distribution}

The null distribution is the least rank statistic of the product of $N$ uniform random variates, each representing an unbiased selection of a P-value $p_{i,x}$ from the empirical distribution of correlation coefficient $r$. $N$ is the number of frames used to calculate a given $P_{x}$.

The least rank statistic was assessed over the number of samples along the LOV with a given $N$ frames. For example, ``complete'' indexes sampling all 9 valid IPTF frames form a set of 180 samples, therefore the corresponding null distribution is calculated as the least statistic of products of 180 sets of 9 uniform random variates.

These distributions were constructed by Monte Carlo with $N_{monte}=500$ trials:

\begin{equation}
    \left\{\prod_{i=1}^{N_x} \left[ u_{i,k}\right]\right\}_{N_{monte}}, u_{i,k}\sim \mathrm{Uniform[0,1]}
\end{equation}

The 95\% confidence interval of the null distribution was calculated for each sample of the LOV shown by the grey band in \cref{fig:JointP}. Each sample in the LOV is matched to its corresponding $N$ number of valid IPTF frames, and the null CI is calculated for each set of $N$ valid frames.

\begin{figure}
\centering
     \includegraphics[width=\linewidth]{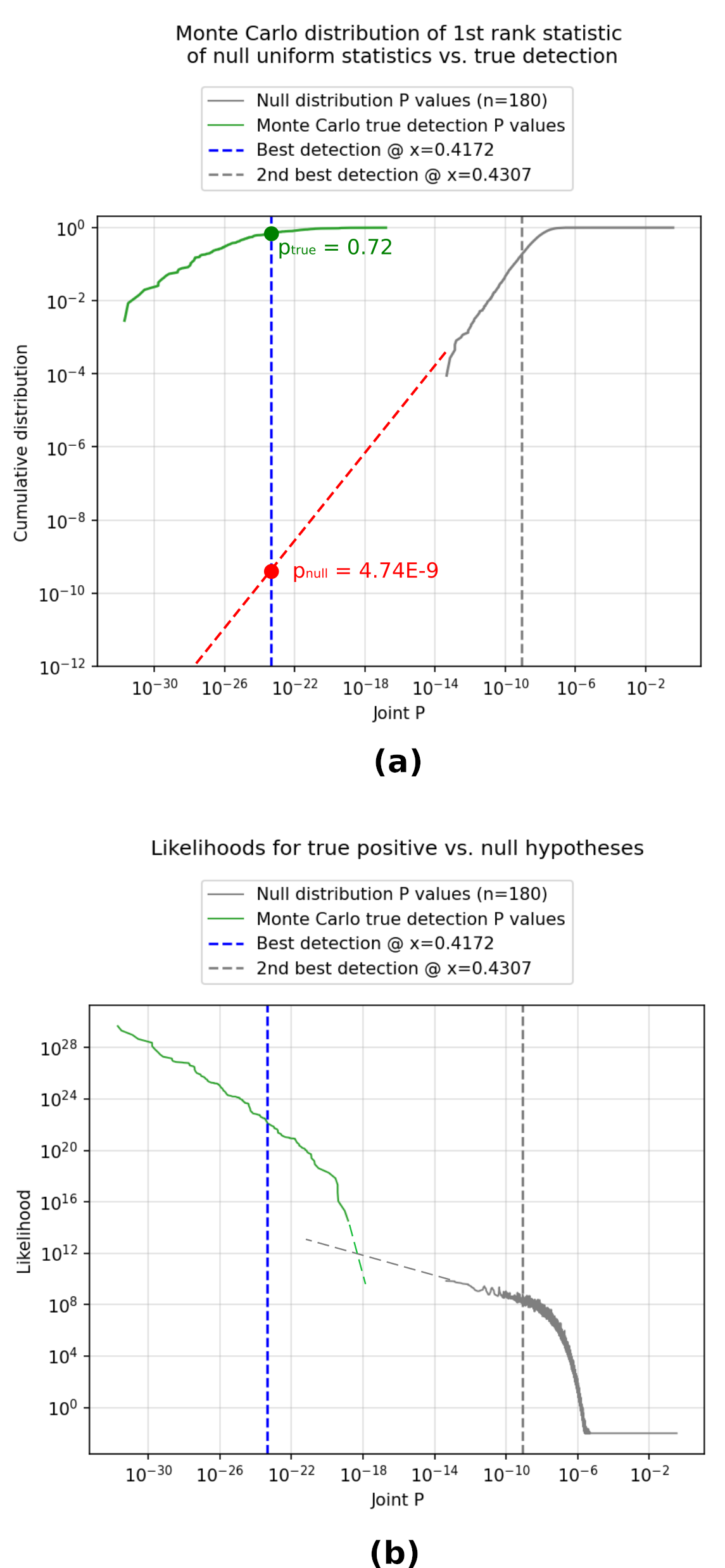}
     \caption{Null vs. true detection hypothesis testing. The joint $P_{x^*}$ of the best detection candidate is shown by the blue dashed line, followed by the second best candidate with the gray dashed line. \textbf{(a)} Log-CDF of the distribution of joint P-values for the null distribution (in gray) and the true detection distribution (in green). For the best detection candidate, the null hypothesis rejected at $P_{null}=4.74\times 10^{-9}$. \textbf{(b)} Log likelihood plots for the true detection (green) vs. null hypothesis (gray). The overlap in likelihood is $P_x\sim 10^{-18}$, while the detection is made at $P_x < 10^{-22}$. In both, the second most significant detection is indicated with a vertical dashed gray line for comparison, showing it to be well within the expected null distribution.}
\label{fig:detection-testing}
\end{figure}

\subsection{True positive distribution}
\label{ssec:true-positive-distribution}

The true positive distribution represents the distribution of joint P-values $P_x$ generated by 2024 YR$_4$ with the addition of CCD noise and light contamination in the image fields. The reference images generated for correlation were scaled and additively composited onto regions of the frames outside of the LOV to create representative samples of the correlation metric caused by true detections of 2024 YR$_4$.

The image pixel values were scaled by zero point fitting of each IPTF image using g- and r-band stellar photometry from the Pan-STARRS 1 DR2 catalog \citep{magnier2020} accessed via Astroquery \citep{2019AJ....157...98G}. Star finding was performed using the DAOPHOT program (\cite{Stetson1987}) implemented in the photutils package (\cite{larry_bradley_2025_14889440}) for stellar sources of magnitude +19 to +13.

Approximately 1000 composited sample detections were generated per image, and these samples were correlated to the reference image to generate distributions of true positive P-values $p^+_{i,x}$ and their products $P^+_x=\prod_{i=1}^{N_x}{p^+_{i,x}}$. The distribution for the ``complete'' set of 9 images is shown in \cref{fig:detection-testing}.

The 95\% confidence interval of the true positive distribution was calculated for each sample of the LOV shown by the green band in \cref{fig:JointP}. Each sample in the LOV is matched to its corresponding \textit{set} of valid IPTF frames, and a corresponding true detection distribution is calculated for each combination of valid frames.

The test demonstrates exceptional power, with the null and true detection distributions for ``complete'' sets of 9 images separated by more than two $95\%$ confidence intervals as shown in \cref{fig:JointP}. Furthermore, the null and true distribution CIs maintain substantial separation at all positions, suggesting 2024 YR$_4$ would be detectable in any location in the LOV.

\subsection{Results}

The joint P-values $P_x$ were calculated for each sample along the LOV and shown in \cref{fig:JointP}, alongside the 95\% confidence intervals for the null and true detection distributions. Only one detection appears in the true detection distribution: a detection of $P_{x^*}=4.452\times10^{-24}$ at $x^*=0.4172$, corresponding to the candidate described in Section \ref{sec:candidate}.

\cref{fig:correlation-boxplots} shows the P-values $p_{i,x^*}$ for the candidate at $x^*=0.4172$. For all but the first image, this candidate's P-values $p_{i,x^*}$ lie within $2\sigma$ of the distributions predicted by 2024 YR$_4$'s expected imaging properties.

\cref{fig:detection-testing} shows the performance of the candidate in the maximum likelihood test. The null hypothesis is rejected at $P_{null} < 10^{-8}$. Conversely, the detection is accepted by the two-sided P test at $P_{true}=1-|0.72-0.5|*2>0.5$.

\begin{figure}
\label{fig:YR4stacks}
\centering
\includegraphics[width=\linewidth,angle=0]{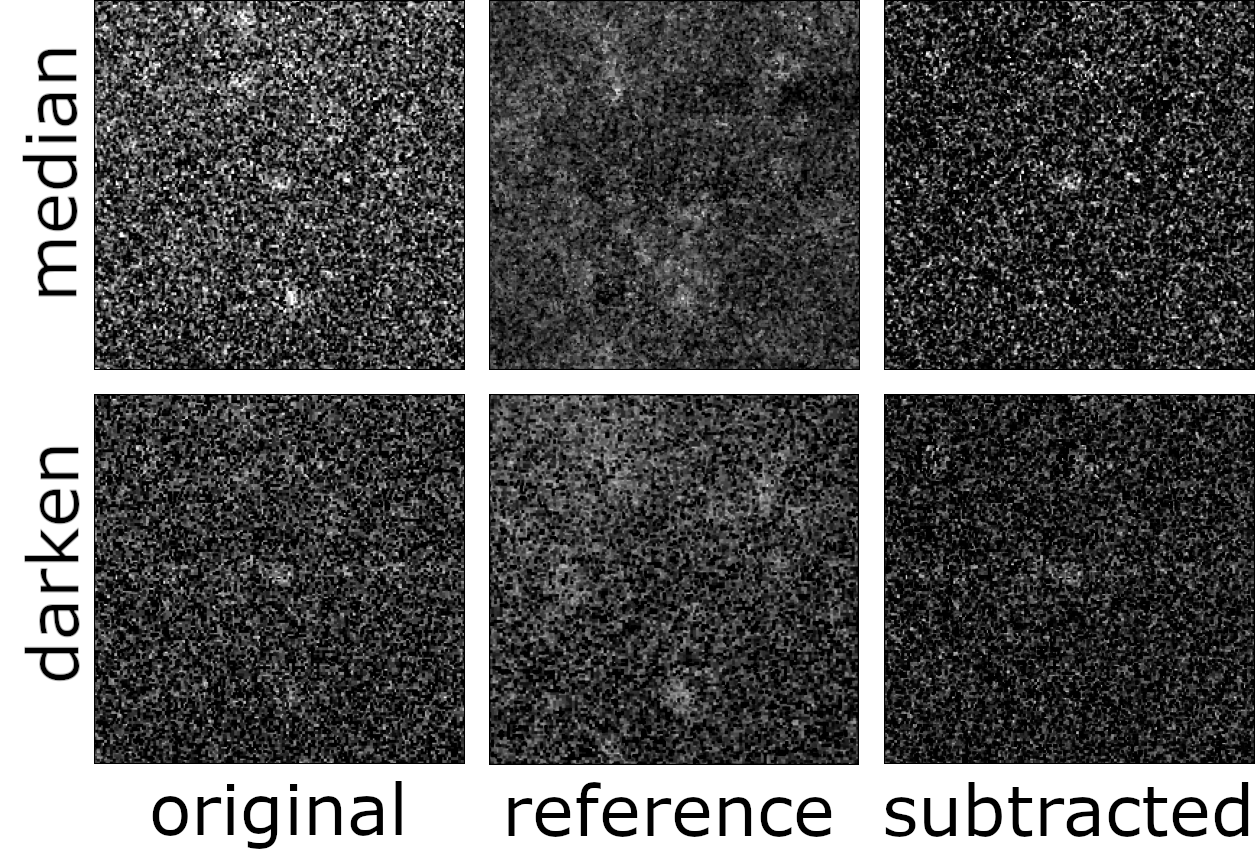}
\caption{An array of various stacks of the candidate detection. (Left column) the IPTF data from all nights centered on the detection's location. (Central column) the ZTF deep reference data corresponding to each image. (Right column) the reference-subtracted IPTF data. (Top row) median stacks of the respective dataset, where the median brightness value for each pixel is preserved. (Bottom row) darken stacks of the respective dataset, where only the darkest pixel of the image series is preserved.
\label{fig:Stacks}}
\end{figure}

We created median and darken stacks centered on the position of the mean detection along the LOV (Figure \ref{fig:YR4stacks}) as a final visual validation of this detection. Although these two methods are standard for extracting signal and assessing the validity of detections in typical asteroid recovery work, in 2024 YR$_4$'s case they are particularly destructive. Both of these stacks require a consistent detection across a large number of images to preserve signal, while 2024 YR$_4$ varies significantly in brightness over the course of the IPTF observations due to both varying observing circumstances at Palomar Observatory and the asteroid's own rotation and gradual approaching of Earth. This consistency is also hampered by its streak length significantly changing between nights, so the final result is that these stacks are particularly free of stellar contamination, but also disproportionately reduce the SNR of the detection relative to random noise due to not taking advantage of any priors other than the expected position on each date.

To further reduce stellar contamination, the same operations were performed on the ZTF reference images, as well as with the reference-subtracted IPTF imagery. The reference imagery shows signals consistently appearing at the same place as the IPTF stacks, corresponding to stars overlapping between images, while these signals are largely removed in the subtracted images as expected.

\section{Results, Orbit and Conclusions}\label{sec:summary}

\begin{deluxetable}{ccccccc}
\label{tab:IPTF measurements}
\tablecaption{IPTF measurements of 2024 YR$_4$ detection}
\tablehead{
\colhead{Date} & \colhead{RA} & \colhead{DEC} & \colhead{RA sig} & \colhead{DEC sig} & \colhead{resid RA} & \colhead{resid DEC}
\\ & \multicolumn{2}{c}{(degrees)} & \multicolumn{4}{c}{(arcsec)}
}
\startdata
2016 08 25.314342 & \multicolumn{6}{c}{\tablenotemark{a}} \\
2016 08 25.350126 & \multicolumn{6}{c}{\tablenotemark{a}} \\
2016 08 25.367710 & 342.135990 & +17.398225 & 1.6 & 2.3 & +0.90 & -0.35 \\
2016 08 25.406723 & 342.156407 & +17.465856 & 1.1 & 1.3 & -0.77 & -0.32 \\
2016 08 28.387770 & 344.421896 & +23.545672 & 1.2 & 1.3 & +0.42 & +0.67 \\
2016 08 28.426011 & 344.453574 & +23.635866 & 0.8 & 1.8 & +0.27 & +0.65 \\
2016 08 31.405845 & 348.033927 & +32.062871 & 0.7 & 1.6 & -0.22 & -2.3 \\
2016 08 31.444627 & 348.089280 & +32.191741 & 0.7 & 0.8 & -0.09 & +0.92 \\
2016 09 03.409365 & 354.527735 & +43.815969 & 0.6 & 1.6 & +0.14 & +1.0 \\
\enddata
\tablenotetext{a}{Too faint/contaminated for meaningful measurement}
\end{deluxetable}

In the previous sections, we identified and described a candidate detection along 2032 Lunar virtual impactor 2024 YR$_4$'s long but thin uncertainty region in archival Intermediate Palomar Transient Factory imagery. When analyzing this against the expected properties of 2024 YR$_4$ in the imagery, we find this detection to be a) statistically significant, b) consistent with 2024 YR$_4$'s expected properties (properties which make 2024 YR$_4$ theoretically detectable across the entire LOV in this imagery) and c) the only significant detection across the entire LOV ($P\approx 5\times 10^{-9}$), with no other signals surpassing the null test.

As a result of these factors, we contend that this detection is indeed most likely that of 2024 YR$_4$, observed over 8 years before its 2024 discovery. This detection, and indeed the source dataset, appears to have gone unnoticed by other initial archival followup projects due to not just the faintness of the detection, but the ongoing unreleased nature of much of the IPTF survey, these images included. We hope that the publication of this paper may draw attention to the merits of IPTF as a valuable archival tool on par with other contemporaneous major surveys (the Catalina Sky Survey, Pan-STARRS, Spacewatch, etc) in case similar precovery work becomes necessary again in the future.

Table \ref{tab:IPTF measurements} catalogs the measured positions, magnitudes, and associated uncertainty for each of the measured positions of 2024 YR$_4$ in the IPTF dataset; each measurement was made independently as a centroid to the modeled streak as predicted for that date, with no effort made to conform to a consistent or expected location. Table \ref{tab:YR4 orbit} provides the resultant calculated orbit solution for 2024 YR$_4$ making use of all available published observations at time of writing in addition to our aforementioned measurements. We used the Find\_Orb software \citep{Gray2022} to calculate the orbit solution, with the published observations of 2024 YR$_4$ acquired in ADES through the Minor Planet Center's Explorer service.\footnote{https://data.minorplanetcenter.net/explorer}

This results in a lunar approach distance of $22001 \pm 49$ km from the Lunicenter, and $277534 \pm 46$ km from the geocenter -- an encounter which will reach a peak magnitude of V=8.0 from the Moon and V=13.5 from Earth. Using this orbit solution, the maximum 1$\sigma$ positional uncertainty 2024 YR$_4$ will achieve is only circa 90 arcseconds, well within the field of view of all but the smallest telescopes.

\begin{deluxetable}{ccc}
\label{tab:YR4 orbit}
\tablecaption{2024 YR$_4$ provisional orbit solution (1$\sigma$ uncertainties)}
\tablehead{
\colhead{Parameter} & \colhead{with 2016} & \colhead{only 2024-2025}
}
\startdata
Epoch & \multicolumn{2}{c}{2029-01-01.0} \\
Semimajor axis (au) & $2.515452983 \pm 8.1 \times10^{-9}$ & $2.515452411 \pm 2.7 \times10^{-6}$ \\
Eccentricity & $0.66086320 \pm 6.1 \times10^{-9}$ & $0.66086308 \pm 4.9 \times10^{-7}$ \\
Inclination ($\deg$) & $3.407822 \pm 5.9 \times10^{-7}$ & $3.407822 \pm 3.7 \times10^{-5}$\\
Long. Node ($\deg$) & $271.370960 \pm 3.7 \times10^{-6}$ & $271.370958 \pm 3.9 \times10^{-5}$ \\
Arg. Peri. ($\deg$) & $134.292566 \pm 3.9 \times10^{-6}$ & $134.292581 \pm 5.3 \times10^{-4}$ \\
Mean Anom. ($\deg$) & $10.624179 \pm 3.5 \times10^{-6}$ & $10.624451 \pm 1.2 \times10^{-3}$ \\
Perihelion (au) & $0.853082668 \pm 1.5 \times10^{-8}$ & $0.853082766 \pm 3.2 \times10^{-7}$ \\
Aphelion (au) & $4.177823299 \pm 2.3 \times10^{-8}$ & $4.177822056 \pm 5.7 \times10^{-6}$ \\
2032 Perilune (km) & $22001 \pm 49$ & $25423 \pm 17195$\\
2032 Perigee (km) & $277534 \pm 46$ & $280982 \pm 16061$\\
Obs. used (rejected) & 490 (20) & 483 (20) \\
Obs. arc (days) & 3135 & 91 \\
Mean residuals & 0.21 arcsec & 0.19 arcsec \\
JWST 2025 net residuals & (-0.003, -0.002) & (-0.005, -0.002) \\
\enddata
\end{deluxetable}

This orbit solution assumes a purely gravitational orbit, although due to its small size 2024 YR$_4$ likely has minor Yarkovsky forces which may slightly affect its encounter with Earth and the Moon. Yarkovsky forces, produced by the unbalanced absorption and emitting of sunlight by a rotating asteroid, are non-negligible over the course of years or decades on small (<100 m-scale) asteroids, having been detected in dozens of near-earth asteroids with multi-year observation arcs. Calculating an orbital solution with a free A2 parameter gives a value of $A2 = +2.71 \pm 3.65\times10^{-12}$ au/day$^2$ (1/r$^2$), an uncertainty roughly an order of magnitude higher than the average A2 value among 86 asteroids with absolute magnitudes $23 < H < 25$ and nonzero A2 values in the JPL Small-Body Database\footnote{https://ssd.jpl.nasa.gov/tools/sbdb\_query.html} ($0.20 \times10^{-12}$ au/day$^2$). Regardless, taking this large Yarkovsky uncertainty at face value results in the nominal Lunar encounter details changing to $19189 \pm 3537$ km -- clearly excluding an impact. Constraining the A2 to be less than $0.56 \times10^{-12}$ au/day$^2$ (the maximum absolute value among the previously mentioned 86 asteroids with recorded A2 parameters) limits the approach distance uncertainty to $\pm \leq522$ km from the nominal (non-gravitational) solution calculated above.

\subsection{Revisiting Non-IPTF imagery}

As we discussed in Section \ref{sec:analysis}, it seems that at least the 2016-08-09 Subaru and 2016-08-11 DECam imagery should be sufficiently deep to detect 2024 YR$_4$ if it was in either field; we additionally identified that the only sections of the LOV missing in both datasets correspond to 2032 perilunes of 20000 through 28000 km and 51000 through 57000 km, which suggested that the true location of 2024 YR$_4$ would be in an orbit within either range. Indeed, the ITPF data detection being located almost exactly at 22000 km is fully consistent with the lack of a detection in either of these datasets.

\subsection{Recovery Prospects}

Due to the critical importance of accurate measurements of 2024 YR$_4$ on account of its highly anticipated and uncertain 2032 encounter with the Earth-Moon system, it is exceptionally important to validate what would be in other circumstances more confident observations. Despite the lengths we have gone to demonstrate the significance and circumstantial likelihood of this detection, it is important that our work is reproducible with time to spare before its major 2032 approach.

To this end, our full orbital solution using the candidate detection is made available in Table \ref{tab:YR4 orbit}, and our specific perilune/perigee predictions are solvable as specific locations along the line of variations at any desired observing period.

The next time 2024 YR$_4$ will get bright enough (regardless of the validity of these detections) will be in 2028, when it will reach a peak magnitude of V=20.8. Without our observations, the 3$\sigma$ uncertainty ellipse is $\pm 37$ arcseconds according to the most recent calculations from JPL Horizons. With our observations, however, we predict a positional uncertainty of only $\pm \leq 0.5$ arcsec around our nominal orbit, a prediction which should be trivial to test due to the relative brightness of the asteroid at the time, and the comparatively small fraction of the current LOV that this prediction makes up. Our hope in this paper's writing is that this will save valuable time from large telescopes to extensively confirm 2024 YR$_4$'s orbit in 2028, in addition to demonstrating advanced image analysis techniques which could be of use if such a high-interest asteroid is discovered in the future. With the framework outlined here, it may now be possible to fully search for, identify, and validate candidates in archival datasets of similar difficulty to those of 2024 YR$_4$'s archival search campaign in a matter of weeks rather than months or years.

\begin{acknowledgments}
We would like to thank Frank Masci and the IPTF science team for their gracious sharing and analysis of unpublished IPTF imagery. We would additionally like to thank Daniel Parrott for his helpful input on photometric analysis methodology and the creation of the Tycho Tracker software which assisted in this paper. This research has made use of the NASA/IPAC Infrared Science Archive, which is funded by the National Aeronautics and Space Administration and operated by the California Institute of Technology. This project used data obtained with the Dark Energy Camera (DECam), which was constructed by the Dark Energy Survey (DES) collaboration. This research is based in part on data collected at the Subaru Telescope, which is operated by the National Astronomical Observatory of Japan. We are honored and grateful for the opportunity of observing the Universe from Maunakea, which has the cultural, historical, and natural significance in Hawaii. This research has made use of "Aladin sky atlas" developed at CDS, Strasbourg Observatory, France.

\end{acknowledgments}

\pagebreak
\bibliography{references}
\bibliographystyle{aasjournal}

\end{document}